\documentclass[pra,amsmath,amsfonts,showkeys,showpacs,preprint]{revtex4}
\usepackage{graphicx}
\RequirePackage{times}
\RequirePackage{courier}
\RequirePackage{mathptm}

\begin{document}

\title{Tracing the bounds on Bell-type inequalities}
\author{Stefan Filipp}
\email{sfilipp@ati.ac.at}
\affiliation{Atominstitut der {\"{O}}sterreichischen Universit{\"{a}}ten,
Stadionallee 2, A-1020 Vienna, Austria}
\author{Karl Svozil}
\email{svozil@tuwien.ac.at}
\homepage{http://tph.tuwien.ac.at/~svozil}
\affiliation{Institut f\"ur Theoretische Physik, University of Technology Vienna,
Wiedner Hauptstra\ss e 8-10/136, A-1040 Vienna, Austria}

\begin{abstract}
Bell-type inequalities and violations thereof reveal the fundamental
differences between standard probability theory and its
quantum counterpart. In the course of previous investigations
ultimate bounds on quantum mechanical violations have been
found. For example, Tsirelson's bound constitutes a global upper limit
for quantum violations of the Clauser-Horne-Shimony-Holt (CHSH) and
the Clauser-Horne (CH) inequalities.
Here we investigate a method for calculating the precise quantum bounds on arbitrary
Bell-type
inequalities by solving the eigenvalue problem for the operator
associated with each Bell-type inequality. Thereby, we use the min-max
principle to calculate the norm of these self-adjoint operators from
the maximal eigenvalue yielding the upper bound for a particular set
of measurement parameters. The eigenvectors corresponding to
the maximal eigenvalues provide the quantum state for which a
Bell-type inequality is maximally violated.
\end{abstract}

\maketitle

\section{Introduction}

One of the most puzzling features of quantum mechanics is the
violation of so-called Bell-type inequalities representing a
cornerstone of our present understanding of quantum probability theory
\cite{peres93}. As pointed out by John Bell \cite{bell64} such a
violation, as predicted by quantum mechanics, requires a radical reconsideration of basic physical
principles like the assumption of local realism. However, Bell-type inequalities have already a long tradition
dating back to George Boole's work on ``conditions of possible
experience'' \cite{boole54, boole62}, dealing with the question of
necessary and sufficient conditions on probabilities of logically
interconnected events \footnote{We are therefore tempted to
  use the term ``Boole-Bell-type inequalities'', but to be in line
  with current terminology we use just ``Bell-type inequalities'' instead.}.
Take for example the statements: ``The probability of rain in V\"axj\"o
is about $80\%$'' and ``The probability
of rain in Vienna is $90\%$''. Nobody would
believe  that the joint probability of rain in both places could be
just $10\%$ --- the claim that the joint probability is very much lower than the single
probabilities is apparently counterintuitive. The question remains:
Which numbers could be considered reasonable and consistent?
Boole's requirements on the (joint) probabilities
are expressed by certain equations or inequalities relating those (joint) probabilities.

Since Bell's investigations \cite{bell64,bell66} into bounds on classical
probabilities and their relation to quantum mechanical predictions,
similar inequalities for particular physical setups have been
discussed in great number and detail (see for example Refs.
\cite{clauser69,clauser74,werner-wolf01,zukowskibrukner02}).
Furthermore, violations of Bell-type
inequalities,
as predicted by quantum mechanics,
have been experimentally verified in different areas of physics
\cite{aspect82, weihs98, rowe01, hasegawa03} to a very good degree of accuracy.

However, whereas these bounds are interesting for an inspection of
the violations of classical probabilities by quantum probabilities,
the issue of the validity of quantum probabilities and their
experimental verification is completely different. Recently, Bovino
\emph{et al.} \cite{bovino04} conducted an experiment based on
numerical studies by the current authors \cite{filipp-svozil04} and
triggered by a proposal of Cabello \cite{cabello04} to verify bounds on
quantum probabilities depending on a particular choice of
measurements.

In what follows we shall present analytical as well as numerical
studies on such quantum bounds allowing for further experimental
tests of different kinds of Bell-type inequalities.

\section{Correlation Polytopes}

At first we shall start from a geometrical derivation of bounds on
classical probabilities given by linear inequalities in
terms of correlation polytopes
\cite{froissart81,pitowsky89,pitowsky01a,
  pitowsky94,tsirelson80}.
Considering an arbitrary number of classical events $a_1, a_2,\ldots ,
a_n$ one can assign to each event a certain probability
$p_1,\,p_2,\,\ldots,\,p_n$ and probabilities $p_{12},\ldots$ for the joint
events $a_1 \cap a_2,\ldots$. These probability values can be
collected to form the the components
of a vector $p=(p_1, p_2,\ldots , p_n, p_{12},\ldots )$,  where each
 $p_i,\,p_{ij}\ (i,j = 1,\ldots, n)$ can take values in the interval $[0,1]$.
 Since the events $a_1,\,a_2,\,\ldots$ are assumed to be independent,
 each single probability $p_i$ can also take its extremal value $0$ or
 $1$ and the
 vectors comprising all possible combinations of extremal values ($p_i=0,1$ and $p_{ij}=p_i p_j$) can be regarded
 as rows of a truth table; with the symbols ``$0$'' and ``$1$'' corresponding to \emph{``false''} and
 \emph{``true,''} respectively.

Any classical probability distribution; i.~e., any vector $p$, can be represented as a convex
sum over the extremal probability distributions given by the row
entries of the truth table. It can therefore be regarded as some point
$p \in C$ where $C={\textrm conv} (K)$ is a convex polytope defined by
the set of all points that can be written as a convex sum extending
over all vectors associated with row entries in the truth table. More formally,
\begin{equation}
  {\textrm conv} (K)=\left\{ \sum_{i=1}^{2^n} \lambda_i{\bf x}_i \;
\left| \; \lambda_i\ge 0,\; \sum_{i=1}^{2^n}\lambda_i =1 \right.
\right\}
\end{equation}
with
\begin{equation}
K = \{{\bf x}_1,{\bf x}_2,\ldots ,{\bf x}_{2^n}\} = \left\{ \left.
\big(t_1, t_2,\ldots , t_n, t_xt_y,\ldots \big) \; \right| \; t_i
\in \{0,1\},\; i=1,\ldots ,n \right\}.
\end{equation}
 Here, the terms
$t_xt_y,\ldots$ stand for arbitrary products associated with the joint
propositions which are considered. Exactly what terms are considered
here depends on the particular physical configuration.

In a next step towards the linear inequalities sought one utilizes the
Minkowsky-Weyl representation theorem \cite[p.29]{ziegler94} stating
that every convex polytope in Euclidean real space has a dual
description: either as the convex hull of its extreme points -
in our case the rows of the truth table -  or as the intersection of a
finite number of half-spaces. Each half space can be described by a linear
inequality.
To obtain the inequalities from the vertices one has to solve the
so-called \emph{hull problem}. These inequalities coincide with Boole's
``conditions of possible experience'';
 i.~e., they constitute the
bounds of classical probabilities.
The set of inequalities obtained
is maximal and
complete, as no other system of inequalities exists which
characterizes the correlation polytope completely and exhaustively.

For particular physical setups these inequalities correspond to
Bell-type inequalities. Therefore correlation
polytopes provide a constructive way of finding the entire set of
Bell-type inequalities for a given physical configuration
\cite{pitowsky01,filipp01}, although from a computational complexity
point of view the problem remains intractable \cite{pitowsky91}.

As an example, we consider the derivation of the well known
Clauser-Horne inequality \cite{clauser74}: Given a source emitting pairs of correlated
spin-1/2 particles either in the positive or in the negative $y$-direction,
the spin of both particles can be measured in arbitrary directions  perpendicular to the
propagation direction;
 i.~e., restricted to the $x$--$z$ plane.
Implementing two measurement directions on each
side labeled by the angles $\alpha,\ \beta$ for the particle
propagating in the negative $y$ direction (left hand side) and
$\gamma,\ \delta$ for the particle propagating along the positive
$y$ axis (right hand side), we
obtain the probabilities for measuring ``spin-up'' for each particle
and measurement direction $p_\alpha, p_\beta, p_\gamma, p_\delta$. The
joint probabilities for measuring ``spin-up'' on the left while
measuring ``spin-up'' on the right in coincidence -- but in general with different measurement
directions -- are denoted by $p_{\alpha\gamma},p_{\alpha\delta},\ldots$
The probability distribution vector for this situation is consequently
$p=(p_\alpha,
p_\beta,p_\gamma,p_\delta,p_{\alpha\gamma},p_{\beta\gamma},p_{\beta\delta},p_{\alpha\delta})$
and the truth table (comprising the extremal probabilities) consists of
$2^4=16$ rows by inserting $p_\alpha,\,p_\beta,\,p_\gamma,\,p_\delta \in \{0,1\}$.
The corresponding polytope is eight dimensional.
By solving the hull problem, which for this simple
setup can easily be done,
we obtain inequalities like $0 \leq
p_\alpha,p_{\alpha\gamma}\leq 1$, $p_\alpha + p_\gamma -
p_{\alpha\gamma} \leq 1$; and in the similar manner for $p_\beta,\
p_\gamma,\ p_\delta$. Additionaly, inequalities of the form
\begin{equation}
  \label{eq:chineq}
    -1\leq p_{\alpha\gamma} + p_{\alpha\delta} + p_{\beta\gamma} - p_{\beta\delta} - p_{\alpha} - p_{\delta} \leq 0
\end{equation}
also represent bounds of this correlation polytope.
The inequality (\ref{eq:chineq}),
termed \emph{Clauser-Horne (CH) inequality}, and the inequalities containing all permutations of the parameters,
are violated by quantum theory
for particular choices of the angles and for specific quantum states.
They constitute therefore a demarcation
criteria between classical and non-classical probabilities, such as the ones encountered in quantum theory.

\section{Violation of Bell-type Inequalities}

Similar to the bounds on classical probabilities given by the
Bell-type inequalities, there exist bounds
on quantum probabilities which will be the subject of the following
discussion.
There have been investigations in the analytic aspects of bounds on
quantum probabilities, most prominently by
Tsirelson \cite{tsirelson80,khalfin92} and recently by others in Refs. \cite{pitowsky01a, masanes03, cabello04, filipp-svozil04a}, but also numerical
\cite{filipp-svozil04} and experimental \cite{bovino04} test have been
performed.
The quantum
probabilities do not
violate Bell-type inequalities maximally
\cite{popescu92,mermin95,krenn-svozil98}.
Take, for example, the well known Clauser-Horne-Shimony-Holt (CHSH)
inequality
\footnote{The CHSH-inequality is defined in terms of
  expectation values instead of probabilitiess its equivalent in terms
  of probabilities being the
  Clauser-Horne (CH) inequality.}
\begin{equation}
\label{CHSH}
|E(\alpha,\gamma)+E(\beta,\gamma)+E(\beta,\delta)-E(\alpha,\delta)|\leq 2,
\end{equation}
where $E(\mu,\nu)$ denotes the correlation function for two particle
correlations with possible values in the interval $[-1,1]$ when measuring their spin/polarization in coincidence
along the directions $\mu$ and $\nu$, respectively. The global limit
for a quantum
violation of this inequality is $2\sqrt{2}$ \cite{tsirelson80,landau87};
quantum theory does not allow a higher value, no
matter which state and which measurement directions are chosen.
However, in principle, the four terms on the left hand
side of Eq. (\ref{CHSH})
could be set such that a value of $4$ can be obtained by appropriate
choices of $\pm 1$ for the correlation functions.

Popescu and Rohrlich \cite{popescu92} investigated the case
where  ``physical locality'' is assumed without referring to a
specific physical model (such as quantum mechanics), whether realistic or
not.
In this context, ``physical locality'' means that the
marginal probabilities for measuring an observable on one side should
be independent of the
observable measured on the other side, which is a natural assumption
for a Lorentz invariant theory. The  maximal value
of the left hand side of Eq. (\ref{CHSH}) has been shown to be $4$ as well, which is beyond the
quantum bound $2\sqrt{2}$ and we can conclude that quantum mechanics
does not exploit the whole range of violations possible in a theory
conforming to relativistic causality.
Still, in our opinion, the nagging question remains why
quantum mechanics does not violate the inequality to a higher
degree.

In what follows, we will restrict our attention to the simpler task to
explore the quantum bounds on violations of Bell-type inqualities for
particular given measurement directions and arbitrary
states.
It turns out that the equations for the analytic description of the
quantum bounds can be derived by
solving an eigenvalue problem.
Intuitively it cannot be expected that it is feasible to
achieve a maximal violation of some inequality for any set of
measurements just by choosing a single appropriate state.

The quantum mechanical description of the physical scenario discussed
above involves spin measurements represented
by projection operators
\begin{equation}
  \label{eq:proj}
  F(\theta) = \frac{1}{2}\left(1_2+\vec{\sigma}(\theta)\right),
\end{equation}
with $\vec{\sigma}(\theta)=\left(\begin{array}{cc}\cos\theta &
  \sin\theta \\ \sin\theta & -\cos\theta\end{array}\right)$, $\theta$
denoting the direction of measurement in the $x$--$z$ plane, and $1_2$
standing for the two-dimensional identity matrix.
For an even
more general description we would have to take all possible
two-dimensional projection operators into account, corresponding to
measurements in arbitrary directions. As this generalization is
straightforward and does not lead to any more insight, we will work
with this restricted set of measurements parameterized in Eq.~(\ref{eq:proj}).

$F(\theta)$ acts on one of the two
particles.
This implies that we have to choose a tensor product of two
Hilbert
spaces to represent the state vectors corresponding to possible state
configurations; i.~e., $\mathcal{H} = \mathcal{H}_1 \otimes
\mathcal{H}_2$. The representation of a single-particle measurement in
$\mathcal{H}$ is then
\begin{equation}
\label{singleparticlemeasurement}
  q(\theta) = \frac{1}{2}\left(1_2+\vec{\sigma}(\theta)\right )\otimes
  1_2\quad \mbox{or}\quad  1_2 \otimes \frac{1}{2}\left(1_2+\vec{\sigma}(\theta)\right)
\end{equation}
for a measurement on the particle emitted in the negative $y$-direction
($\mathcal{H}_1$), or in the positive $y$-direction ($\mathcal{H}_2$), respectively. Two-particle measurements are
implemented by applying $F(\theta)$ on both $\mathcal{H}_1$
and $\mathcal{H}_2$; i.~e.,
\begin{equation}
\label{twoparticlemeasurement}
  q(\theta,\theta') = \frac{1}{2}\left(1_2+\vec{\sigma}(\theta)\right)
  \otimes \frac{1}{2}\left(1_2+\vec{\sigma}(\theta')\right),
\end{equation}
corresponding to a measurement of the joint probabilities.
This setup can easily be enlarged to systems comprising more than two
particles by the tensor product of  the appropriate Hilbert spaces,
but for the sake of simplicity we will restrict
ourselves to bipartite systems.

The general method for obtaining the quantum violations of Bell-type
inequalities is then to replace the classical probabilities by
projection operators in
Eqs. (\ref{singleparticlemeasurement},\ref{twoparticlemeasurement}) in
a certain Bell-type inequality
to obtain the \emph{Bell-operator}, which is a sum of projection
operators. In the case of the CH inequality one obtains
\begin{equation}
  -1\leq q(\theta_\alpha,\theta_\gamma) +
  q(\theta_\alpha,\theta_\delta) + q(\theta_\beta,\theta_\gamma) -
  q(\theta_\beta,\theta_\delta) - q(\theta_\alpha) - q(\theta_\delta)
  \leq 0.
\end{equation}
In a second step one calculates the quantum
mechanical expectation values by
\begin{equation}
  \label{eq:expect}
  \langle q(\theta) \rangle = \mbox{Tr}[W q(\theta)],
\end{equation}
where $W$ is a positive definite, Hermitian and normalized density
operator denoting the state of the system. For some $W$ and set of
angles $\{\theta_\alpha,\theta_\beta,\theta_\gamma,\theta_\delta\}$
one obtains a violation of a classical inequality.

In general the Bell-operators can be written in the form
\begin{equation}
  \label{eq:Belloperator}
  O =  \sum_{i_1,i_2,\ldots,i_N} b_{i_1i_2\ldots i_N} P_{i_1}\otimes
    P_{i_2} \otimes \ldots P_{i_N},
\end{equation}
with real valued coefficients $b_{i_1i_2\ldots i_N}$.
Here $N$ is the number of particles involved and the $P_i$ are either
projection operators denoting a measurement on particle $i$ or the
identity when no measurement is performed on the $i$-th particle. Since  $(A \otimes B)^\dagger = A^\dagger
\otimes B^\dagger = A \otimes B$ and $(A+B)^\dagger = A^\dagger +
B^\dagger = (A+B)$ for arbitrary selfadjoint operators $A,B$, the
Bell-operator $O$ is also self-adjoint with real eigenvalues. However,
the eigenvalues of $O$ cannot be deduced from the eigenvalues of the
constituents $P_{i_1}\otimes
    P_{i_2} \otimes \ldots P_{i_n}$ in the sum in
    Eq. (\ref{eq:Belloperator}) since these are not commuting in
    general and therefore are not diagonalizable simultaneously.

    One can make use of the \emph{min-max principle} \cite[\S
    90]{halmos74}, stating that the bound of a
    self-adjoint operator is equal to the maximum of the absolute
    values of its eigenvalues. Thus, the problem of finding the
    maximal violation possible for a particular choice of measurements
    can be solved via an
    eigenvalue problem. The maximal eigenvalue corresponds to the
    maximal violation and the associated eigenstates are the
    multi-partite states which yield a maximum violation of the
    classical bounds under the given experimental (parameter) setup 
    \footnote{Non-degenerate eigenstates are always representable by
      one-dimensional subspaces and thus are pure, the exception being
      the possibility of a mixing between degenerate eigenstates
      \cite{braunstein92}.}.

For a demonstration of the method let us start with the trivial setup
of two particles measured along a single (but not necessarily
identical) direction on either side.
The vertices are $(p_1,p_2,p_{12}=p_1p_2)$ for $p_1, p_2 \in \{0,1\}$ and thus
$(0,  0,  0)$,
$(0,  1,  0)$,
$(1,  0,  0)$,
$(1,  1,  1)$;
the corresponding face (Bell-type) inequalities of the polytope spanned by the four vertices
are given by
$p_{12} \le p_2$,
$0\le p_{12}\le 1$, and $p_1+p_2-p_{12}\le 1$.

The classical probabilities have to be substituted by the quantum ones;
i.e.,
\begin{equation}
\begin{array}{lll}
p_1 &\rightarrow& q_1 (\theta ) =
{\frac{1}{2}}\left[1_2 + {\bf \sigma}( \theta )\right] \otimes 1_2,
\\
p_2 &\rightarrow& q_2 (\theta ) =
1_2 \otimes {\frac{1}{2}}\left[1_2 + {\bf \sigma}( \theta )\right],
\\
p_{12}&\rightarrow& q_{12} (\theta ,\theta ') =
{\frac{1}{2}}\left[1_2 + {\bf \sigma}( \theta )\right]
\otimes
{\frac{1}{2}}\left[1_2 + {\bf \sigma}( \theta ')\right].
\end{array}
\label{2004-qbounds-e2}
\end{equation}

It follows that the self-adjoint transformation corresponding to the classical Bell-type inequality
($p_1+p_2-p_{12}\le 1$) is given by
\begin{equation}
O_{11}(0,\theta) = q_1(0)+q_2(\theta)-q_{12} (0, \theta )
=\left(
\begin{array}{cccccccc}
   1 & 0 & 0 & 0 \cr 0 & 1 & 0 & 0 \cr 0 & 0 & {\cos^2
     \frac{\theta }{2}} & \frac{\sin \theta }
   {2} \cr 0 & 0 & \frac{\sin \theta }{2} & {\sin^2  \frac{\theta
     }{2}} \cr
 \end{array}\right).
\end{equation}

The eigenvalues of $O_{11}$ are $0$ and $1$, irrespective of $\theta$,
the maximal value of $O_{11}$ predicted by the min-max principle
 does not exceed the classical bound 1.

Now we shall enumerate analytical quantum bounds
for the more interesting cases comprising two or three distinct
measurement directions on either side yielding the
quantum equivalents of the Clauser-Horne (CH) inequality, as well as
of the inequalities discussed in \cite{pitowsky01,collins-gisin03,sliwa03}.

For two measurement directions per side,
we obtain the operator $O_{22}$ based on the CH-inequality [Eq.~(\ref{eq:chineq})]
upon substitution of the classical
 probabilities by projection operators:
\begin{equation}
O_{22}(\alpha,\beta,\gamma,\delta)=  q_{13}(\alpha,\gamma) +
q_{14}(\alpha,\delta) + q_{23}(\beta,\gamma) \nonumber - q_{24}(\beta,\delta)- q_{1}(\alpha) - q_{3}(\gamma).
\label{2004-qbounds-e4}
\end{equation}

The eigenvalues of the self-adjoint transformation in
(\ref{2004-qbounds-e4}) are
\begin{equation}
  \label{eq:2004-qbounds-evo22}
  \lambda_{1,2,3,4}(\alpha,\beta,\gamma,\delta )
=
\frac{1}{2}\big(\pm\sqrt{1\pm\sin(\alpha -\beta )\sin(\gamma -\delta )}-1\big),
\end{equation}
yielding the maximal bound
$\|O_{22} \|= \max_{i=1,2,3,4} \lambda_i$.
The eigenstates
corresponding to maximal violating eigenstates are maximally
entangled for general measurement angles lying in the $x$--$z$-plane  \cite{filipp-svozil04a}.

The numerical simulation of the bounds of the CH-inequality
is based
on the generation of arbitrary bipartite density matrices $W$; i.~e., $4
\times 4$ Hermitian positive matrices with trace equal to one.
Since one can write a Hermitian positive matrix $W$ as the square of a
self-adjoint matrix, $W=B^2$.
The
normalized matrix $W'=W/\mbox{Tr[W]}$ can thus be explicitly parameterized by $16$ parameters
$b_1,b_2,\ldots ,b_{16}\in {\mathbf R}$; i.~e.,
\begin{equation}
    \label{e-2003-qpoly-2}
    W'=\frac{1}{\sum_{i=1}^4 b_i^2 + 2 \sum_{j=5}^{16}
    b_j^2}\left(\begin{array}{cccc} b_1 & b_5 + {\rm i}b_6 & b_{11} +
    {\rm i}b_{12} & b_{15} + {\rm i}b_{16}\\ b_5 - {\rm i}b_6 & b_2 &
    b_7 + {\rm i}b_8 & b_{13} + {\rm i}b_{14}\\ b_{11} - {\rm i}b_{12}
    & b_7 - {\rm i}b_8 & b_3 & b_9 + {\rm i}b_{10}\\ b_{15} - {\rm
    i}b_{16} & b_{13} - {\rm i}b_{14} & b_9 - {\rm i}b_{10} &b_4
      \end{array}\right)^2 .
\end{equation}

\begin{figure}[htbp]
  \centering
  \includegraphics[width=80mm]{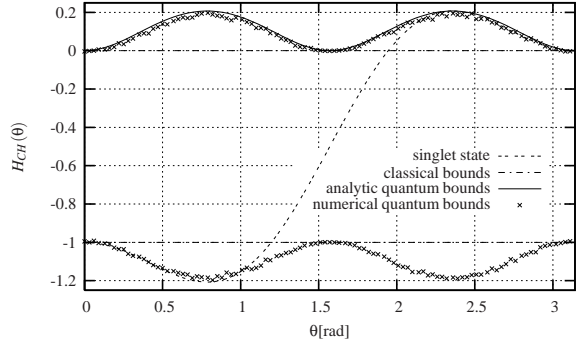}
  \caption{Numerical simulation of the bounds of the CHSH-inequality}
  \label{fig:numerical}
\end{figure}

For a particular choice of projection operators, one can then generate
random states $W'$ in order to find the maximal violation possible for the
current set of projection operators. In Figure \ref{fig:numerical}, both the
analytic and the numerical bounds are depicted for measurement directions $\alpha=0,\
\beta=2\theta,\ \gamma=\theta$ and $\delta=3\theta$ dependent on a
single parameter $\theta \in [0,\pi]$.
In addition, the well-known maximal
violation for the singlet-state at $\theta=\pi/4$ and $\theta=3\pi/4$
is drawn.

\begin{figure}[htbp]
  \centering
  \includegraphics[width=90mm]{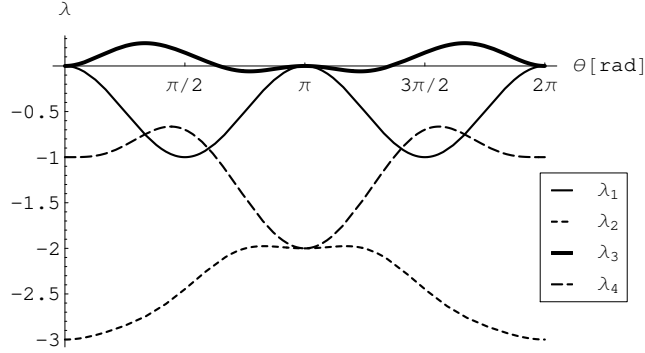}
  \caption{Eigenvalues of $O_{33}$ in dependence of the relative angle $\theta$.}
  \label{fig:2004-qbounds-f1}
\end{figure}

The extension to \emph{three} measurement operators for each particle merely
yields one additional non-equivalent inequality (with respect to
symmetries) \cite{collins-gisin03,sliwa03}
\begin{equation}
I_{33}=p_{14} + p_{15} + p_{16} + p_{24} + p_{25} - p_{26} + p_{34} - p_{35}
- p_{1} - 2 p_{4} - p_{5} \leq 0
\end{equation}
 among the 684 inequalities \cite{pitowsky01} representing the
 faces of the associated classical correlation polytope.
The associated operator for symmetric
measurement directions is given by
\begin{equation}
\begin{array}{lll}
&O_{33}(0,\theta,2\theta,0,\theta,2\theta)= q_{14}(0,0) + q_{15}(0,\theta) + q_{16}(0,2\theta) + q_{24}(\theta,0) +
q_{25}(\theta,\theta) - q_{26}(\theta,2\theta) +\\
&\quad  + q_{34}(2\theta,\theta)- q_{35}(2\theta,\theta)-q_{1}(0) - 2 q_{4}(0) - q_{5}(\theta) \\
&\quad =\frac{1}{4}\left(
\begin{array}{cccc}
-4\sin^2\theta & 0 & 0 & 0\\
0 & -5-2\cos\theta - 3\cos 2\theta + 2\cos 3\theta &
4\cos^2\frac{\theta}{2} & 2\sin\theta + 3 \sin 2\theta - 2 \sin
3\theta\\
0 & 4\cos^2\frac{\theta}{2} & -2(3+\cos 2\theta) & - 2\sin\theta \\
0 &  2\sin\theta + 3 \sin 2\theta - 2 \sin 3\theta & - 2\sin\theta &
2\sin^2\frac{\theta}{2}\cos^2\frac{\theta}{2}(4\cos\theta -3)
\end{array}\right),
\end{array}
\label{2004-qbounds-e5}
\end{equation}

in the Bell basis
$\{|\phi^+ \rangle,|\psi^+ \rangle,|\psi^- \rangle,|\phi^- \rangle\}
\equiv \{(1,0,0,0)^T,(0,1,0,0)^T,(0,0,1,0)^T,(0,0,0,1)^T\}$
with
$|\psi^\pm \rangle = 1/\sqrt{2}(|01 \rangle \pm |10 \rangle)$ and
$|\phi^\pm \rangle = 1/\sqrt{2}(|00 \rangle \pm |11 \rangle)$.
In this basis, $O_{33}$ can be decomposed into a direct sum of a
one-dimensional and a three-dimensional matrix $O_{33} = o_1 \oplus
o_3$, thus simplifying the calculations of the real eigenvalues.
By using the Cardano method \cite{cocolicchio00}, these can be calculated to be
\begin{eqnarray}
\lambda_1 &=& -\sin^2\theta,\nonumber\\
\lambda_2 &=& -2 \sqrt{|u|}\cos\frac{\xi}{3}-\frac{b}{3},\nonumber\\
\lambda_{3,4}& =& \sqrt{|u(x)|}\Big[\cos\frac{\xi}{3} \pm
\sin\frac{\xi}{3}\Big]-\frac{b}{3}.
\label{2004-qbounds-o33ev}
\end{eqnarray}
Here, $u=1/9(3 c-b^2)$ and $\cos\xi = \frac{1}{54}\big(9 b c -2 b^3 -
27 d\big)/\big(u\sqrt{|u|}\big)$ where
$b = -\mbox{Tr}\, o_3,\ c = 1/2\Big(\mbox{Tr}\,^2 o_3 -
\mbox{Tr}\, o_3^2 \Big),\ d = -\det o_3$.
(For convenience we have omitted the dependencies on $\theta$.)
In Figure \ref{fig:2004-qbounds-f1},
the eigenvalues are plotted  as functions of the parameter $\theta$.
The maximum violation of $1/4$ is obtained for $\theta=\pi/3$ with the
eigenvector corresponding to $\lambda_3$
\begin{equation}
  |\Psi_{\rm max } \rangle=\frac{\sqrt{3}}{2}|\phi^- \rangle+\frac{1}{2}|\psi^+ \rangle.
\label{2004-qbounds-pmax33}
\end{equation}

$|\Psi_{\rm max } \rangle$ is maximally entangled, but in contrast to
the CH-inequality, this is in general not the case for
eigenstates corresponding to the maximal eigenvalue at $\theta \neq
\pi/3$.

\section{Relation to Experiments}

The analytical quantum bound of the CH-operator has been
enumerated by Cabello \cite{cabello04} as well as by the current authors
\cite{filipp-svozil04a} and experimentally verified by
Bovino \emph{et al.} \cite{bovino04} using polarization-entangled
photon pairs.
 The ansatz of Cabello for the experimental realization made use
of the fact that the eigenstates leading to maximal violations are
maximally entangled. Thus when applying a unitary transformation $U(\theta)$ of the form
$$U(\theta) = \left(\begin{array}{cc}\sin\theta & -\cos\theta\\\cos\theta & \sin\theta
  \end{array}\right)$$
onto an initial state
$|{\psi^-}\rangle=\frac{1}{\sqrt{2}}(|{01}\rangle-|{10}\rangle)$,
one obtains all maximally violating states for different $\theta$
values $|{\psi(\theta)}\rangle = U(\theta)\otimes
1_2 |{\psi^-}\rangle$.

However, in the case of $O_{33}$,  this scheme has to be extended, since
the maximal violating states are not maximally entangled in general.
Such states cannot be created from maximal entangled initial states
by a local unitary
operation $U_{2\times 2} \in SU(2) \otimes SU(2)$,
since such a factorized transformation does not change the degree of
entanglement.
To obtain states constituting the quantum bounds, one has to apply unitary transformations
$U_4 \in SU(4)$ to the initial state comprising also non-local operations which  cannot be written as a
tensor product of two unitary single-particle operators. 

A simplification for an experimental verification of the quantum bounds of
Bell-type inequalities is due to the fact that maximal violating
states are pure.
Therefore, it is sufficient to generate initial
states with variable degree of entanglement. Utilizing the
Schmidt-decomposition, which is always possible for a bipartite state,
one can write any pure state in the form $|{\psi} \rangle= \sum_k \lambda_k
|{k_1}\rangle \otimes |{k_2}\rangle$ where $|{k_j}\rangle$ are orthonormal basis
states for particle $1$ and $2$, respectively, and $\sum_k \lambda_k^2 =
1$. The weights of the $\lambda_k$'s are a measure of the degree
of entanglement comprising the special cases where $\lambda_1 =
\lambda_2 = 1/\sqrt{2}$ for a maximally entangled state and $\lambda_1
= 0,\ \lambda_2 = 1$ (or vice versa) for a separable state. Having a
source producing such states in a particular basis one can obtain all
other pure states by applying a local unitary operation
$U_{2\times 2} \in SU(2) \times SU(2)$. Appropriate photon sources have been
suggested for example by White \emph{et al.} \cite{White:1999} and Barbieri
\emph{et al.} \cite{barbieri04} and could therefore be used to trace the bounds
on arbitrary bipartite Bell-type inequalities in the same manner as in
the experiment of Bovino \emph{et al.} \cite{bovino04}.

\section{Conclusion}

In conclusion we have shown how to obtain analytically the quantum
bounds on Bell-type inequalities for a particular choice of
measurement operators. We have also presented a numerical simulation
for obtaining these bounds for the CH-inequality.
We have provided a quantitative
analysis and derived the exact quantum bounds for bipartite
inequalities involving two or three measurements per site. The
generalization to an arbitrary number of measurement parameters is
straightforward as  the dimensionality of the eigenvalue problem
remains constant. For more than two particles the dimension of the
matrix associated with a Bell-type operator increases exponentially.
However, one may conjecture that such matrices can
be decomposed into a direct sum of lower dimensional matrices.

In the context of this conference we also believe that the analytic expressions of the quantum bounds
could serve as consistency criteria of  mathematical
models proposed to show that a violation of Bell-type inequalities
does not necessarily imply the absence of a possible local-realistic
theory from the logical point of view. It is claimed that violations can be achieved
without abandoning a local and realistic position assuming for example
time-dependencies of
the random parameters \cite{hess-philipp01a},  or ``chameleon'' effects
\cite{accardi01}.
Still, any appropriate
model has to be in accordance with quantum mechanics not only
qualitatively, but also quantitatively, and hence should reproduce also the ``fine structure'' of
the quantum bounds as discussed above.

Finally, although there is no theoretical evidence for a stronger-than-quantum
violation whatsoever, its mere possibility justifies the sampling
of the fine structure of the quantum bounds from the experimental as well as the theoretical
point of view in order to understand and verify the restriction imposed by
quantum theory.

\section{Acknowledgments}
S. F. acknowledges the support of the Austrian
Science Foundation, Project Nr. 1514, and the support of
Prof. H. Rauch rendering investigations in such theoretical aspects
of quantum mechanics possible.

\bibliographystyle{aipprocl}


\end{document}